\documentclass[twocolumn, preprintnumbers, 
endnote,nofootinbib,prl,9pt, superscriptaddress]{revtex4}
\usepackage{soul}
\usepackage{epsfig,subfigure}

\usepackage{epstopdf}

\usepackage[utf8]{inputenc}
\usepackage{graphicx}
\usepackage{amssymb}
\usepackage{amsxtra}
\usepackage{amsmath}
\usepackage{booktabs,multirow,tabularx}
\usepackage{slashed}
\usepackage{float}
\usepackage{placeins}
\usepackage{rotating}
\usepackage{lscape}
\usepackage{color}
\usepackage{hyperref}
\usepackage{soul}

\usepackage[Q=yes,pverb-linebreak=no]{examplep}

\DeclareGraphicsExtensions{.eps}
\graphicspath{{./}}

\newcommand{\lsim}{\lesssim}
\newcommand{\gsim}{\gtrsim}
\newcommand{\eps}{\varepsilon}

\newcommand{\ord}[1]{\mathcal{O}{(#1)}}
\newcommand{\eq}[1]{Eq.~(\ref{#1})}

\newcommand{\DBL}{\Delta (B-L)}

\newcommand{\lam}{\lambda}

\def\beq{\begin{equation}}
\def\bea{\begin{eqnarray}}
\def\eeq{\end{equation}}
\def\eea{\end{eqnarray}}
\def\beqnl{\begin{align}}
\def\endal{\end{align}}

\definecolor{red1}{cmyk}{0,1,1,0.1}
\definecolor{blue1}{cmyk}{1,0,0,0}

\DeclareFontFamily{U}{cbgreek}{}
\DeclareFontShape{U}{cbgreek}{m}{n}{
        <-6>    grmn0500
        <6-7>   grmn0600
        <7-8>   grmn0700
        <8-9>   grmn0800
        <9-10>  grmn0900
        <10-12> grmn1000
        <12-17> grmn1200
        <17->   grmn1728
      }{}
\DeclareFontShape{U}{cbgreek}{bx}{n}{
        <-6>    grxn0500
        <6-7>   grxn0600
        <7-8>   grxn0700
        <8-9>   grxn0800
        <9-10>  grxn0900
        <10-12> grxn1000
        <12-17> grxn1200
        <17->   grxn1728
      }{}

\makeatletter
\newcommand{\normalorbold}{%
  \ifnum\pdf@strcmp{\math@version}{bold}=\z@ bx\else m\fi
}
\makeatother

\begin{document}

\date{\today}

\title{\boldmath Higgs Troika for Baryon Asymmetry}

\author{Hooman Davoudiasl\footnote{email: hooman@bnl.gov}
}

\affiliation{Physics Department, Brookhaven National Laboratory,
Upton, New York 11973, USA}

\author{Ian M. Lewis\footnote{email: ian.lewis@ku.edu}
}

\author{Matthew Sullivan\footnote{email: mattsullivan14916@ku.edu}
}

\affiliation{Department of Physics and Astronomy, University of Kansas, Lawrence, Kansas, 66045 USA}

\begin{abstract}

To explain the baryon asymmetry of the Universe, we extend the
Standard Model (SM) with two additional Higgs doublets with small vacuum expectation values.  The additional Higgs fields interact with SM fermions through complex Yukawa couplings, leading to new sources of CP violation.  We propose a simple flavor model with $\mathcal{O}(1)$ or less Yukawa couplings for quarks and charged leptons, consistent with current flavor constraints.  To generate neutrino masses and the baryon asymmetry, right-handed neutrinos in the $\sim 0.1-10$~TeV range couple to the ``Higgs Troika.''  The new Higgs doublet masses are at or above the TeV scale, allowing for asymmetric decays into SM lepton doublets and right-handed neutrinos.  The asymmetry in lepton doublets is then processed into a baryon asymmetry, similar to leptogenesis.   Since the masses of the new fields could be near the TeV scale, there is potentially a rich high energy collider phenomenology, including observable deviations in the 125~GeV Higgs decay into muons and taus, as well as detectable low energy signals such as the electron EDM or $\mu\rightarrow e\gamma$.  Hence, this is in principle a testable model for generation of baryon asymmetry. 
 
\end{abstract}

\maketitle

\section{Introduction\label{sec:intro}}

The Standard Model (SM) of particle physics has been successful in explaining a wide range of 
phenomena and remains valid after many years of experimental verification.  Yet, 
the SM leaves important 
fundamental questions unanswered.  Among these, the origin of dark matter and the source of 
the baryon asymmetry of the universe (BAU) - both of great importance to our understanding of cosmology and matter - remain open.  While dark matter may reside in an 
entirely secluded ``dark sector,'' it is reasonable to expect that the physics underlying the BAU must have direct and perhaps significant interactions with the SM and is part of the ``visible sector.'' 

In this work, we propose to extend the SM content with two additional Higgs fields, copies of the 
SM Higgs, with small vacuum expectation values (vevs).  While these fields will have a marginal role in electroweak symmetry breaking (EWSB), they could have significant complex-valued 
couplings to the SM fermions and provide 
new sources of CP violation.  We will show that this setup is then capable of accommodating a 
baryogenesis mechanism, as long as the new Higgs masses are at or above the TeV scale.  

Our basic mechanism is in spirit similar to leptogenesis \cite{Fukugita:1986hr}, however we do not require heavy right-handed neutrinos $\nu_R$ far above the 
weak scale, whose role will be assigned to the new Higgs scalars here.  We will choose right-handed neutrino masses in the $\sim$ 0.1-10~TeV range to implement 
our scenario.  Our proposal is a minimal realization of ``neutrinogenesis''~\cite{Dick:1999je,Murayama:2002je}. The SM extended to include a ``Higgs Troika'' can then explain the origin of visible matter and the masses of fundamental particles.  This setup can be potentially testable at colliders in the future, perhaps even at high luminosity LHC (HL-LHC) with $\ord{\text{ab}^{-1}}$ levels of data, expected to be available in the coming years.    

Next we will briefly outline our mechanism and  
describe the main ingredients and assumptions underlying 
our proposal.  We will then illustrate the mechanism in a benchmark realization of the model and provide some quantitative estimates.  A brief discussion of the benchmark collider phenomenology 
will also be given, in order to highlight some of the key features of the possible signals.  For some related ideas in a different 
context, see Refs.~\cite{Davoudiasl:2011aa,Gu:2006dc}.

\section{The Baryogenesis Mechanism \label{sec:bg}}

Here, we briefly describe the general features of the baryogenesis mechanism.  Let us denote the Higgs fields by $H_a$ with masses $m_a$, $a=1,2,3$.  We will identify $H_1$ as the observed (``SM'') Higgs with $m_1 \approx 125$~GeV: $H_1 \leftrightarrow H_{\rm SM}$.  This implies that $H_1$ has the same Yukawa interactions as the SM Higgs and generates the known masses of fermions.  Also, it is implicitly assumed that new interactions of $H_1$ with other scalars are  sufficiently small to avoid significant deviations from the SM predictions for the main Higgs production and decay modes.  To make contact with potential experimental searches, we will generally assume that $m_{2,3} \sim1$~TeV (this mass scale may also originate from the physics underlying the SM Higgs sector, though we will not dwell on this point further).  

In order to generate a baryon asymmetry, we need an asymmetry in the decays of $H_i$ and $H_i^*$ into SM fermions which will lead to an asymmetry in the number density of the SM fermions.  The total decay rates of $H_i$ and $H_i^*$ are equal by CPT.  However, the partial decay rates of $H_i$ and $H_i^*$ do not have to be equal.  Hence, to generate an asymmetry $\eps$, we need at least two different decay channels for the new scalars.  We will specify those interactions later, however, here we will only mention that one of the channels is $\bar L \nu_R$ (which we will refer to as ``neutrinos''), with $L$ a lepton doublet in the SM. 

 Although $\nu_R$ is a lepton, it is a gauge singlet.  Hence, {\it sphalerons} will not operate on $\nu_R$.  The relevant non-zero $\Delta(B-L)$ is then for quark and lepton doublets; where $B$ is baryon number and $L$ is lepton number.  That is, sphalerons will not act on an asymmetry in $\nu_R$ nor alter the baryon asymmetry generated via the lepton doublets~\cite{Dick:1999je,Murayama:2002je}. The other channel is provided by coupling to SM charged fermions.  The asymmetry requires a non-zero CP violating phase to remain in the interference of tree and 1-loop diagrams; this in turn requires at least two Higgs scalars that couple to leptons and quarks, implying that we at least need $H_2$.  Below, we will illustrate why we also need $H_3$, on general grounds.  However, briefly put, since the $H_1$ mass is at the EW scale and not larger than the reheat temperature, it could efficiently mediate processes that washout the baryon asymmetry.  Hence, we need three Higgs doublets, i.e. a ``Higgs Troika.''

Let us denote a  typical Higgs coupling to $\bar L\, \nu_R$ by $\lambda_a^\nu$ and 
to charged fermions by $\lambda_a^f$.  For concreteness and simplicity, we will assume that the asymmetry is dominated by the $f$ intermediate fermion, but the width of $H_2$ is set by decays into the fermion $f'$.  This assumption implies $\lambda_2^{f'}$ is the dominant Yukawa coupling of $H_2$; we consider this a fairly generic assumption.  The asymmetry, as will be discussed later in more detail, is typically then given by  
\beq
\eps \sim \frac{\lambda_1^\nu \lambda_1^f \lambda_2^\nu\lambda_2^{f}}
{8 \pi \, (\lambda_2^{f'})^2}.
\label{eps-est}
\eeq    
On general grounds, an asymmetry parameter of order $\eps\gsim 10^{-9}$ is needed to generate the BAU \cite{Tanabashi:2018oca}
\beq
\frac{n_B}{s} \approx 9\times 10^{-11}\,,
\label{BAU-obs}
\eeq  
where $n_B$ is the baryon number density and $s$ is the entropy density.  

Here, we note that the success of our baryogenesis scenario requires that 
$2\to 2$ processes 
$F f \to L \nu_R$, where $F$ is an $SU(2)_L$ doublet and $f$ an $SU(2)_L$ singlet,
through the interactions of $H_1$ should not washout the generated $\DBL$.  This 
requirement should be maintained down to a temperature of $T_* \sim 100$~GeV, below 
which EWSB takes place.  Above that temperature all SM fields, except the Higgs, can be assumed to be massless.  
Hence, the rate for washout at $T=T_*$ is roughly given by 
$\Gamma_* \sim (\lambda_1^\nu \lambda_1^f)^2 T_*$.  Requiring that $\Gamma_* \lsim H(T_*)$, 
where $H(T) \approx g_*^{1/2} T^2/M_P$ is the Hubble scale,  
one finds 
\beq
\lambda_1^\nu \lambda_1^f \lsim \left(\frac{g_*^{1/2} T_*}{M_P}\right)^{1/2}\,,
\label{lambda1}
\eeq
where $g_*\sim 100$ denotes relativistic degrees of freedom and 
$M_P\approx 1.2 \times 10^{19}$~GeV is the Planck mass.  The above yields $\lambda_1^\nu \lambda_1^f \lsim 10^{-8}$.  

To generate the asymmetry parameter in Eq.~(\ref{eps-est}), there are three interesting cases for the relative strengths of the different $H_2$ couplings:
\begin{enumerate}
\item First consider $\lambda_2^\nu\lambda_2^{f} \ll (\lambda_2^{f'})^2$.  The washout bound then implies $\eps \ll 4 \times 10^{-10}$, which 
suggests that baryogenesis is not feasible.  
\item Next, $\lambda_2^{\nu}\ll \lambda_2^f\sim \lambda_2^{f'}$.  The washout bound together with $\eps\gtrsim 10^{-9}$ then implies that
\begin{eqnarray}
\lambda_2^{\nu}\gtrsim 2.8\,\lambda_2^f.
\end{eqnarray}
This bound is inconsistent with our starting assumption implying that a baryon asymmetry cannot be generated with this hierarchy of couplings.  The results are similar for $\lambda_2^f\ll \lambda_2^\nu\sim \lambda_2^{f'}$.
\item Finally, assume all couplings are similar $\lambda_2^f\sim\lambda_2^\nu\sim\lambda_2^{f'}$.  The washout bound implies that $\eps\lesssim 4\times 10^{-10}$.  That is, baryogenesis is still not feasible.
\end{enumerate}
This conclusion leads us to require a third Higgs doublet 
field $H_3$, to avoid reliance on a light $H_1$, whose interactions are constrained\footnote{See Refs.~\cite{Hambye:2016sby,Hambye:2017elz} for another minimal realization of BAU generation via Higgs decays that relies on highly degenerate Majorana neutrino masses.}.

Successful baryogenesis requires that the reheat temperature $T_{rh}$, here assumed to be set by  the decay of a modulus $\Phi$, is low enough that $2\to2$ washout 
processes mediated by $H_a$, $a=2,3$, are also inefficient.  Note however that we need $T_{rh}>100$~GeV to have effective electroweak sphaleron processes that are required to provide a source of baryon number violation.   Since $T_{rh} < m_a$, for out of equilibrium decay of $H_a$, the rate for this process is of order $(\lambda_a^f \lambda_a^\nu)^2 T_{rh}^5/m_a^4$.  This production rate must be less than the Hubble scale $H(T_{rh})$.  We thus obtain 
\beq
(\lambda_a^f \lambda_a^\nu)^2 \lsim  \frac{g_*^{1/2} m_a^4}{M_P T_{rh}^3}
\quad ; \quad \text{(no washout).}
\label{Trh}
\eeq
For $T_{rh}\gsim 100$~GeV and $m_a \sim 1$~TeV, we roughly obtain 
$\lambda_a^f \lambda_a^\nu \lsim 10^{-6}$.  Note that this constraint is much 
less stringent than the one obtained for $H_1$ before, which could in principle allow 
a large enough value of $\eps$, using $H_2$ and $H_3$.   

As a proof of concept that such a low reheat is possible, consider the decay of a modulus $\Phi$.  At an early time, the Universe was in a matter dominated era due to the oscillation of $\Phi$.  These oscillations are damped via the $\Phi$ decays and the Universe enters a radiation dominated era.  The reheat temperature $T_{rh}$ of the radiation dominated era is estimated as $H(T_{rh})\sim \Gamma_\Phi$, where $\Gamma_\Phi$ is the total width of the modulus and $H(T_{rh})$ is the Hubble parameter at the reheat temperature. Assuming that $\Phi$ couples to a Higgs doublet via the interaction $(\Phi/\Lambda)D_\mu H_i^\dagger D^\mu H_i$, the decay width of $\Phi$ is then
\begin{eqnarray}
\Gamma(\Phi)\sim \frac{1}{32\pi}\frac{m_\Phi^3}{\Lambda^2}\label{eq:PhiDec}
\end{eqnarray}   
Then the reheat temperature is estimated as
\begin{eqnarray}
T_{rh}\sim\left(\frac{1}{32\,\pi\,g_*^{1/2}}\frac{m_\Phi^3\,M_P}{\Lambda^2}\right)^{1/2}.\label{eq:Trh}
\end{eqnarray}
For a modulus mass $m_\Phi\sim 100$~TeV and cut off scale $3\times 10^{10}$~TeV, we find a reheat temperature of $T_{rh}\sim 100$~GeV.

Before going further, we will point out an issue that will inform our benchmark model parameter 
choices later in this work.  The light neutrino masses $m_\nu$ are generated via 
integrating out the heavy Majorana neutrinos to create the Weinberg operator:
\begin{eqnarray}
({\lambda_{1}^\nu})^2\frac{(LH_1)^2}{m_R}.
\end{eqnarray}
The expression for $m_\nu$ is given by  
\begin{eqnarray}
m_\nu\sim \frac{(\lambda_1^\nu)^2}{2}\frac{v^2_{EW}}{m_R}\,,
\label{mnu}
\end{eqnarray}  
where $v_{EW} = 246$~GeV.  \eq{lambda1} for $f=t$ (the top quark, with $\lambda_t \approx 1$), leads to $\lambda_1^\nu \lsim 10^{-8}$. Assuming $m_\nu\sim 0.1$~eV,  we then find
\beq
m_R \lsim 10~\text{MeV}.
\label{mR}
\eeq

The above bound on $m_R$ is in conflict with our assumption that the new physics, including 
$\nu_R$, is at or somewhat above the weak scale.  We will address this question later, showing that 
certain choices of parameters in the minimal model can avoid this conflict.  Briefly put, 
the resolution will amount to the minimal assumption that there are only two massive SM neutrinos 
around $\sim 0.1$~eV and that the third eigenstate could be much lighter and nearly massless, given 
the current state of knowledge of neutrino parameters.

Let us now briefly outline how the above Troika of Higgs fields can 
lead to a viable baryogenesis mechanism.  We will assume that a population of $(H_3,H_3^*)$ is produced non-thermally, such as  
through the modulus $\Phi$ decay in the 
early Universe, but no significant population of $(H_2, H_2^*)$ is present; this could be a result of preferential $\Phi$ decay (see for example, Ref.~\cite{Davoudiasl:2010am}, for such a possibility in a different model).  The CP violating decays of $H_3$ then generate a non-zero 
$B-L$ number from $H_3 \to \bar L\nu_R$.  The 
asymmetry $\DBL$ can get processed into a $\Delta B$ and $\Delta L$ through electroweak  sphaleron processes that are active at temperatures $T \gsim 100$~GeV.

\section{The general model \label{sec:model}}

Here we will introduce the general structure of the model that could realize the above baryogenesis mechanism.  We will not write down all the possible interactions that the model could contain and 
only specify those that are key for our discussions.  To generate the BAU in the manner described above, let us consider the following Yukawa interactions for the Higgs Troika
\beq
\lambda_a^u \tilde H_a^* \bar Q \,u + \lambda_a^d H_a^* \bar Q\, d + 
\lambda_a^\nu \tilde H_a^* \bar L \,\nu_R + \lambda_a^\ell H_a^* \bar L \,\ell\,,
\label{Yukawa}
\eeq
where $a$ labels the Higgs scalars, but the implicit fermion generation indices have been suppressed.  
In the above, $\lambda_a^u$ and $\lambda_a^d$ denote couplings associated with 
the up-type and down-type quarks; the corresponding couplings to neutrinos and charged leptons are 
denoted by $\lambda_a^\nu$ and $\lambda_a^\ell$. 

\begin{figure}[tb]
\includegraphics[width=0.45\textwidth,clip]{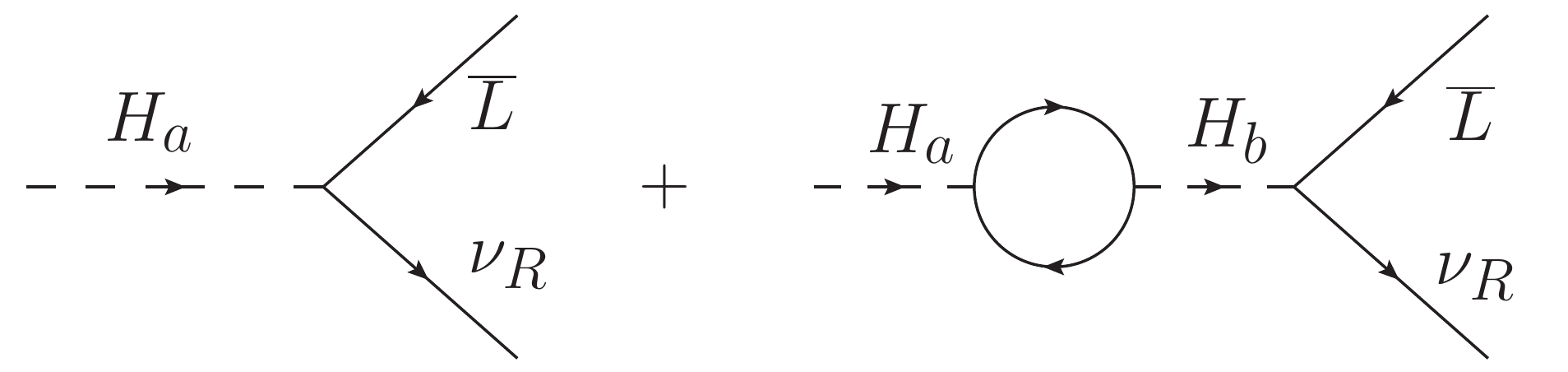}
\caption{Representative tree level and one-loop diagrams that can give rise to a lepton asymmetry.}
\label{fig:eps}
\end{figure}

Let us focus on $a=2,3$.  The $\DBL$ asymmetry $\eps$ produced in the out-of-equilibrium decay of $H_a$ is then given by 
\beq
\eps \equiv \frac{\Gamma(H_a\to \bar L \nu_R) - \Gamma(H_a^*\to \bar \nu_R L)}{2 \Gamma(H_a)},
\label{eps-def}
\eeq  
where $\Gamma(H_a)$ is the total width of $H_a$.  The above is obtained from the interference of the tree and loop-level diagrams in Fig.\ref{fig:eps}.  A second ``triangle'' loop diagram is in general present in our model, but the ``bubble'' loop diagram gets enhanced if the heavy Higgs states $H_2$ and $H_3$ are degenerate in mass (similar arguments apply to heavy right-handed neutrinos in leptogenesis; see for 
example Refs.~\cite{Flanz:1994yx,Pilaftsis:1997jf,Pilaftsis:1998pd}).  

In this work, 
we will consider the case where the heavy Higgs bosons 
$H_2$ and $H_3$ are mildly degenerate and hence we can mostly ignore the ``triangle'' 
contribution to the asymmetry in Fig.\ref{fig:eps}.  Additionally, since our calculation of the baryon asymmetry is an order of magnitude estimation, this approximation is sufficient to show the viability of our mechanism.  For complete expressions see Ref.~\cite{Dick:1999je}.   This assumption simplifies the treatment and also leads to potentially richer collider  phenomenology, as both $H_2$ and $H_3$ can, in principle, be   experimentally accessed.  In this case, the model 
will yield more easily to direct experimental verification.

From Eq.~(\ref{Yukawa}), we find 
\bea
\eps =\frac{1}{8\pi}\sum_{b\neq a}\frac{m_a^2}{m_b^2-m_a^2}\frac{\sum_{f=\ell,u,d}
N_{c,f}\,{\rm Im}\left({\rm Tr}_{ba}^{\nu}{\rm Tr}^{f*}_{ba}\right)}
{\sum_{f=\ell,u,d,\nu} N_{c,f}{\rm Tr}_{aa}^f}
\label{eps}
\eea
where
\bea
{\rm Tr}_{ba}^f&=&{\rm Tr}\left[{\lambda_b^f}^\dagger\lambda_a^f\right]\label{Trf},\\
{\rm Tr}_{ba}^\nu&=&{\rm Tr}\left[{\lambda_b^\nu}^\dagger\lambda_a^\nu(1-m_R^2/m_a^2)^2\right],\label{Trnu}
\eea
$m_R$ are the masses of the right-handed neutrinos, and $N_{c,f}=1,3$ for $f=\text{lepton, quark}$, respectively.  The trace in Eqs.~(\ref{Trf},\ref{Trnu}) is over the fermion generations and we are working in the basis in which $m_R$ are diagonal mass matrices.  Note that $m_f = 0$ during the epoch where $\eps$ is set, since electroweak symmetry is not broken at that point.  As these are traces, the asymmetry parameter in Eq.~(\ref{eps}) can be evaluated in any fermion basis.

In order to find the baryon asymmetry $\Delta B$, we need to find the relationship between $\DBL$ 
and $\Delta B$ in our model, at $T\gsim 100$~GeV.  We note that our setup is that of the SM augmented by two new doublets, however the new doublets are assumed heavy compared to $T_{rh}$ and the relevant field content is that of the SM only.  Also, the processes involving $\nu_R$ are decoupled at these temperatures, as a requirement in our scenario.  Using the results of   
Ref.~\cite{Harvey:1990qw} we then have
\beq
\Delta B = \frac{28}{79}\, \DBL.
\label{BBL}
\eeq 

We will focus on $H_3$ decays, and only consider an intermediate $H_2$.  That is, $a=3$ and $b=2$ in Eq.~(\ref{eps}).  
Given that $\DBL$ is generated through the decay of $H_3$, to calculate the BAU we need to consider the initial energy 
density $\rho_3$ of $H_3$ compared to the radiation energy density $\rho_R$. In our setup, the radiation is made up of all the SM states, including $H_1$.  The decays of $H_3$ contribute to reheating the Universe.  Since $E_3 n_3 \leq \rho_R$, with $n_3$ the number density of $H_3$ and $E_3$ the energy of $H_3$ from decays of the modulus $\Phi$ in Eq.~(\ref{eq:PhiDec}), the ratio 
\beq
r \equiv \frac{E_3 n_3}{\rho_R}, 
\label{r}
\eeq
satisfies $r \leq 1$.

We have $\rho_R = (\pi^2/30) g_* T^4$, where $g_*$ is the number of relativistic degrees of freedom, which is $g_* = 106.75$ in the SM.  The $B-L$ abundance is then given by 
\beq
\frac{n_{B-L}}{s} =  \frac{3\, r \,T_{rh} \, \eps}{4\, E_3},
\label{B-L-abund}
\eeq
where the entropy density $s = (2 \pi^2/45) g_* T^3$. Using \eq{BBL} we then obtain for the BAU 
\beq
\frac{n_B}{s} = \frac{21}{79} \left(\frac{r \,T_{rh} \, \eps}{E_3}\right).
\label{BAU}
\eeq
As shown  in Eq.~(\ref{eq:Trh}), for a modulus $m_\Phi\sim 100$~TeV, we can accommodate a reheat temperature of $T_{rh}\sim 100$~GeV.  Then the energy of $H_3$ is $E_3\sim 50$~TeV and $T_{rh}/E_3\sim 2\times 10^{-3}$.  Hence, for $r\lesssim 1$, one then requires $\eps \gsim 2\times 10^{-7}$ to generate the BAU. 

To show that all the conditions on washout and BAU can be met, we choose a parameter point for proof of concept:\\
\begin{gather}
\begin{tabular}{ll}
$m_\Phi=100$~TeV & $m_3=1.5$~TeV\\
$\lambda_2^\ell\sim 1$ & $\lambda_2^\nu\sim 2\times10^{-6}$\\
$\lambda_3^\ell\sim 1.4\times10^{-3}\quad\quad\quad$ & $\lambda_3^\nu\sim 1.4\times10^{-3}$
\end{tabular}\label{eq:Param1}
\end{gather}
where $m_\Phi$ is the mass of the modulus that generates the $H_3$ population and reheats the Universe, see Eqs.~(\ref{eq:PhiDec},\ref{eq:Trh}).  Additionally, we will assume $H_2$ and $H_3$ only couple to charged leptons and neutrinos.  This assumption and these values for the Yukawas will be motivated in the flavor model presented in the next section.   First, the values of $\lambda_{2,3}^\ell$ and $\lambda_{2,3}^\nu$ satisfy the wash-out condition of Eq.~(\ref{Trh}): $|\lambda_a^\ell\lambda_a^\nu|\lesssim 2.1\times10^{-6}$.  Second, we must check that we can produce the correct BAU, i.e. $\eps\gsim 2\times10^{-7}$.  From Eq.~(\ref{eps}) we have
\begin{eqnarray}
\eps =\frac{1}{8\pi}\frac{m_3^2}{m_2^2-m_3^2}\frac{|\lambda_2^\ell \lambda_2^\nu \lambda_3^\ell \lambda_3^\nu|\sin\phi}{|\lambda_3^\ell|^2+|\lambda_3^\nu|^2}\sim \frac{4\times10^{-8}}{(m_2/m_3)^2-1},\label{eq:eps1}
\end{eqnarray}
where $\phi$ is a generic CP phase.
For $10\%$ level degeneracy between the masses $m_2\sim 1.1 m_3$ and order one phases, the asymmetry parameter is $\eps\sim 2\times10^{-7}$ and the BAU can be generated.  This level of degeneracy is consistent with our assumption that the diagram in Fig.~\ref{fig:eps} is the dominant contribution to the calculation of $\varepsilon$.

Finally, since $H_3$ decays are populating the baryon asymmetry, we must check that they decay much quicker than they annihilate away. The annihilation rate of $H_3$ is calculated by weighting the annihilation cross section, $\sigma_{ann}(H_3)$, by the number density, $n_3$, of $H_3$ in the early Universe:
\begin{eqnarray}
\Gamma_{ann}(H_3)&=&\sigma_{ann}(H_3)n_3.
\end{eqnarray}
We assume $H_3$ couples to one lepton generation with strength $1.4\times 10^{-3}$ and the quartic couplings with the other Higgses are of order $0.1$.  We implement our model into \texttt{FeynRules}~\cite{Alloul:2013bka} and output model files for \texttt{MadGraph5\_aMC@NLO}~\cite{Alwall:2014hca}.  The $H_3$ are produced via the decay of a modulus with mass of $100$~TeV.  Hence, they have energies of 50~TeV.  The annihilation cross sections for 50~TeV $H_3$ into fermions, gauge bosons, and scalars are found to be
\begin{eqnarray}
\sigma_{ann}({\rm fermions})&=&0.43~{\rm fb},\,\nonumber\\
\sigma_{ann}({\rm gauge\,bosons})&=&0.24~{\rm fb},\,\nonumber\\
\sigma_{ann}({\rm scalars})&=&0.17~{\rm fb}
\end{eqnarray}
and the total annihilation rate of $H_3$ is
\begin{eqnarray}
\Gamma_{ann}\lesssim 1.5\times10^{-7}~{\rm GeV},
\end{eqnarray}
assuming $r\leq 1$ and using Eq.~(\ref{r}).  The boosted decay rate into one lepton generation with $H_3$ mass $1.5$~TeV is
\begin{eqnarray}
\Gamma(H_3\rightarrow SM)=\frac{(\lambda_3^\ell)^2}{16\pi}\frac{m_3}{\gamma}=1.8\times10^{-6}~{\rm GeV},
\end{eqnarray}
where $\gamma=E_3/m_3$ is the boost factor.  Hence, the annihilation rate is an order of magnitude smaller than the decay rate, showing the viability of our scenario.

The couplings of $H_3$ are highly constrained by the combination of washout condition and the creation of a large $\eps$ in Eq.~(\ref{eq:eps1}).  To maximize Eq.~(\ref{eq:eps1}), we need to $\lambda_3^\ell \sim \lambda_3^\nu$.  Together with the washout condition this creates the bound $\lambda_3^{\ell,\nu}\lesssim 1.4\times 10^{-3}$.  The couplings of $H_2$ are not so tightly constrained and can be generically larger than those of $H_3$.  Hence, the above model could easily lend itself to collider searches.  In  particular, if the couplings to quarks are not too small, one of  
the heavy Higgs states could be produced at the LHC or a future hadron collider.  Also, depending on the size of the parameters, the rate for decay into charged leptons, a final state with missing energy, or displaced vertices may be large enough 
to enable clean searches.  While there are too many possibilities to consider, we 
will examine a sample benchmark flavor structure choice and describe the main aspects of 
its phenomenology, below.

\section{A benchmark model of flavor \label{sec:flavor}}
Now we give a more complete model of flavor to show our leptogenesis mechanism can work in realistic scenarios.  We introduce three Higgs doublets $\Phi_1,\Phi_2,\Phi_3$.  All three scalar doublets obtain vacuum expectation values $\langle \Phi_i\rangle = v_i/\sqrt{2}$.  The Higgs doublets $\Phi_{2,3}$ and lepton doublets $L$ are odd under a $Z_2$ symmetry while all other fields are even.  The Yukawa interactions are then
\begin{eqnarray}
y_1^u\tilde{\Phi}_1^* \bar{Q}u+y_1^d \Phi_1^*\bar{Q}d+\sum_{b=2,3}y_b^\nu\tilde{\Phi}_b^*\bar{L}\nu_R+y_b^\ell\Phi_b^* \bar{L}\ell.\label{Yukawa1}
\end{eqnarray}
The organizing principle for the charged fermion flavor is that the largest Yukawa coupling for quarks and charged leptons should be order one.  To get the top mass correctly, we need $v_1\approx v_{EW}=246$~GeV.  If there are no fine cancellations there must also be a hierarchy between the vevs $v_{1,2}$ in order to have order one Yukawa for $\tau$.  We would then need $v_2\sim2.5$~GeV for $\lam_a^\tau\sim 1$, while the top quark mass is obtained from the coupling 
to $H_1$ with a Yukawa coupling near unity.  
Since neutrino masses and mixing are rather special and do not follow the patterns of quarks or charged leptons, we do not impose any requirement on their Yukawa couplings.  In principle all neutrinos can get their masses from $H_2$ and one could assume $v_3 \to 0$, though this is not strictly necessary.

Next, we will illustrate how the necessary vev hierarchy can be easily obtained.  Allowing for soft-breaking of the $Z_2$, the relevant terms in the scalar potential are
\begin{gather}
-\mu^2\Phi_1^\dagger\Phi_1+m_2^2\Phi_2^\dagger\Phi_2+m_3^2\Phi^\dagger_3\Phi_3\nonumber\\
-\left(\mu_{12}^2\Phi^\dagger_1\Phi_2+\mu_{13}^2 \Phi^\dagger_1\Phi_3+{\rm h.c}\right)+\lambda(\Phi_1^\dagger\Phi_1)^2+\cdots,
\end{gather}
where $\cdots$ are additional quartics that are not important to this story\footnote{Assuming our hierarchy of scales, we have explicitly checked that the additional quartics make only non-leading contributions to our mechanism.}.  In principle, there is also a $\mu_{23}^2\Phi^\dagger_2\Phi_3$ term, but it can be removed via a rotation of $\Phi_{2,3}$.  This rotation leaves the picture unchanged since $\Phi_{2,3}$ have the same quantum numbers.

For the baryogenesis mechanism to work, we assume the fields $\Phi_{2,3}$ are heavy with $m_2,m_3\sim1$~TeV.  In order for the $Z_2$ breaking to be soft and below 
the highest scales in our theory, we will additionally assume $\mu_{12},\mu_{13}\ll m_{2,3}$.  Once $\Phi_1$ obtains a vev, it induces tadpole terms for $\Phi_{2,3}$.  These tadpoles in turn induce vevs in $\Phi_2$ and $\Phi_3$:
\begin{eqnarray}
v_2 \approx v_1\frac{\mu^2_{12}}{m^2_2}\ll v_1\,{\rm and}\, v_3 \approx v_1\frac{\mu^2_{13}}{m^2_3}\ll v_1.\label{eq:vevs}
\end{eqnarray}
Hence, the tadpole terms give a seesaw where the smallness of $v_{2,3}$ comes from the larger values of the masses $m_{2,3}$.  For $m_2\sim 1$~TeV, $v_2 \sim 2.5$~GeV 
can be generated with $\mu_{12}\sim 100$~GeV.

In order to relate this model to the baryogenesis mechanism, we need to rotate the gauge eigenbasis $\Phi_{1,2,3}$ into the doublet mass eigenbasis $H_{1,2,3}$.  To order $\mu^2/m^2_{2,3}$, this can be accomplished via the rotation
\begin{eqnarray}
\begin{pmatrix} H_1\\H_2\\H_3\end{pmatrix}\approx\begin{pmatrix} 1 & \mu_{12}^2/m_2^2&\mu_{13}^2/m_3^2\\ -\mu_{12}^2/m_2^2&1&0\\-\mu_{13}^3/m_3^2&0&1\end{pmatrix}\begin{pmatrix}\Phi_1\\\Phi_2\\\Phi_3\end{pmatrix}.\label{eqn:rotation}
\end{eqnarray}
From Eq.~(\ref{eq:vevs}), this is precisely the rotation into the Higgs basis such that $\langle H_1\rangle = v_{EW}/\sqrt{2}$ and $\langle H_2\rangle =\langle H_3\rangle = 0$, where $v_{EW}^2=v_1^2+v_2^2+v_3^2\approx v_1^2$.  The Higgs potential is then
\begin{eqnarray}
-\mu^2H_1^\dagger H_1+m_2^2H_2^\dagger H_2+m_3^2 H_3^\dagger H_3+\lambda (H_1^\dagger H_1)^2+\cdots
\end{eqnarray}
That is, $H_{2,3}$ are the doublet mass eigenstates appearing in Eqs.~(\ref{Yukawa}-\ref{eps}), as we desired.  The Yukawas in Eq.~(\ref{Yukawa}) are related to those in Eq.~(\ref{Yukawa1}) via
\begin{eqnarray}
\begin{array}{ll}
\lambda_{1}^{u,d}\approx y_{1}^{u,d},\,& \lambda_{2,3}^{u,d}\approx \displaystyle y_1^{u,d}v_{2,3}/v_{EW},\\
\displaystyle\lambda_{1}^{\ell}\approx y_2^{\ell}v_2/v_{EW},&\lambda_{2,3}^{\ell}\approx y_{2,3}^\ell,\\
 \lambda_{1}^\nu\approx (y_2^\nu v_2+y_3^\nu v_3)/v_{EW},\quad&\lambda_{2,3}^\nu\approx y_{2,3}^\nu,
\end{array}\label{YukawaTrans}
\end{eqnarray}
where we have used $v_2\gg v_3$.  

As discussed previously, there is some tension between washout conditions, having heavy right-handed neutrinos, and generating the light neutrino masses $m_\nu\sim0.1~{\rm eV}$.  We now discuss the neutrino parameters needed to alleviate this tension, supplementing the parameter choices in Eq.~(\ref{eq:Param1}).  
Other choices of parameters may be possible, yet it suffices for our purposes to provide a particular, but not very special, 
realization of our model.  Let $m_{Ri}$, with 
$i=1,2,3$ denote the masses of the three right-handed neutrinos $\nu_{Ri}$.  We will assume that 
$m_{R1,2}\gg m_{2,3}$ and hence the $H_{2,3}$ would not decay into them, while $\nu_{R3}$ 
is light compared to $H_{2,3}$.  We will choose $m_{R3}\sim 100$~GeV and $m_{R1,2}\sim 10$~TeV.  This means that the generation of asymmetry will result from the decay of 
$H_3 \to \nu_{R3}\, \bar L$, with the other channel provided by decay into charged leptons.

Let us take the simplified limit of $v_3\to 0$, corresponding to $\mu_{13}\to 0$, for illustrative purposes.  We will also take the minimal approach of providing two neutrino masses of 
$\ord{0.1\text{eV}}$, with the third state very light or massless, as allowed by all available data.  
For $m_{R1,2}\sim 10$~TeV and $m_\nu\sim 0.1$~eV, \eq{mnu} requires $\lambda_1^\nu \sim 10^{-5}$ which from \eq{YukawaTrans} yields $y_2^\nu \sim 10^{-3}$.  From the 
discussion leading to \eq{Trh}, one could easily determine that washout mediated by $H_{2,3}$ 
could be avoided if we have 
\beq
|\lam_{2,3}^\ell \,\lam_{2,3}^\nu| \lesssim 2.1\times10^{-6}, 
\label{H23washout}
\eeq
where $\ell = e,\mu,\tau$.  
The lepton number violating processes that we would like to avoid correspond to the final 
states $\bar L \,\nu_R$ and its Hermitian conjugate.  Note that for $T_{rh} \gsim 100$~GeV, production of $\nu_{R1,2}$ would be 
severely Boltzmann suppressed, since $m_{R1,2}/T_{rh}\sim 100$.  For final states including 
$\nu_{R3}$, processes mediated by $H_1$ can be decoupled, since $\nu_{R3}$ is not 
required to have substantial coupling to $H_1$ if we only need two mass eigenstates with 
$m_\nu \sim 0.1$~eV.  Hence, we only need to make sure that processes mediated 
via $H_{2,3}$ that lead to a $\nu_{R3}$ in the final state are sufficiently suppressed, corresponding 
to condition Eq. (\ref{H23washout}).

Note that since we require $\lam_2^\tau\sim 1$, suppression of washout mediated by $H_2$ requires 
$\lam_2^{\nu_{R3}} \lsim 10^{-6}$, where the superscript is specified for clarity.  
This, according to \eq{YukawaTrans}, would lead to 
$\lambda_1^{\nu_{R3}}\lsim 10^{-8}$, which is too small to generate $m_\nu\sim 0.1$~eV.  However, as mentioned before, this is consistent with the phenomenologically viable possibility of having one very light neutrino.  Also, as shown in the discussion around Eq.~(\ref{eq:eps1}), these parameter choices are consistent with our baryogenesis mechanism.

We also note that the above sample parameter space leads to $\nu_{R3} \to L \,H_1$ being a typical decay mode of $\nu_{R3}$, as will be shown later when we discuss collider signatures of our model.  
The rate for this decay is estimated to be $\Gamma(\nu_{R3}\to L \,H_1) \sim 
(32\pi)^{-1} \,|\lambda_1^{\nu_{R3}}|^2 m_{R3} \lsim 10^{-16}~\text{GeV}$ which leads to  decays {\it after} EWSB when sphalerons are decoupled.  Hence, $\nu_{R3}$ decays would not interfere with our baryogenesis mechanism.  We then conclude that baryogenesis can be successful in our scenario with the above choice of parameters, as a concrete example.

\section{Low Energy Searches \label{sec:lowenergy}}

We must ensure that the values of Yukawas and CP phases deduced from our benchmark flavor model are consistent with low energy observables such as electric dipole moments.  Additionally, to have a non-zero asymmetry parameter $\eps$, the Yukawas of $H_{2,3}$ must be misaligned.  This misalignment necessarily gives rise to flavor changes that can be searched for.  Up until now we have discussed the couplings of the Higgs doublets.  However, after EWSB we should consider the mass eigenstates in the broken phase, i.e. neutral scalars, neutral pseudoscalars, and charged scalars.  Since the mass eigenstates can mix, their couplings are different than the doublets.  For simplicity of notation, in this section and the next we keep the notation $\lambda_a^f$ for the Yukawas after EWSB.
\begin{itemize}
\item Electric Dipole Moments (EDMs):  The nucleon EDM gets contributions from complex Yukawa couplings of the Higgs fields as well as a $\theta$ term in the QCD Lagrangian.  Assuming a sufficiently small $\theta$ (the usual ``strong CP problem''), we will consider the contribution of  
$H_2$, since the Yukawa coupling of $H_3$ to light quarks is relatively suppressed by a factor 
of $v_3/v_2 \ll 1$ in our flavor model.  See for example Ref.~\cite{Bertuzzo:2015ada} for bounds on a neutrinophilic Higgs doublet.

As we are mostly interested in illustrating that typical values of parameters in our scenario lead to successful baryogenesis and acceptable phenomenology, we will only present order-of-magnitude estimates here.  Since the coupling of quarks to $H_2$ is suppressed by $v_2/v_{EW}$ in our 
flavor model, we find that the 2-loop ``Barr-Zee'' diagrams \cite{Bjorken:1977vt,Barr:1990vd} 
are more important that the 1-loop 
process.   Here, the coupling of $H_2$ to photons is dominated by the $\tau$ loop, which 
couples to $H_2$ with strength $\lam_2^{\tau} \sim 1$, whereas the coupling of the top quark to 
$H_2$ is $\lam_2^t\sim 0.01$.  However, the top mass is about two orders of magnitude larger, which compensates for the suppressed coupling.  Given that these two contributions are roughly similar, we will only use the $\tau$ contribution for our estimate of the effect.  

The 2-loop contribution of $H_2$ (for $v_3 \to 0$ we can ignore $H_3$) to the EDM of a light quark $q$ can then be estimated by 
\beq
d_q \sim \frac{e^3 \, \lam_2^\tau \,\lam_2^q\, m_\tau \,\sin \omega}{(16 \pi^2)^2\, m_a^2}\,,
\label{dq}
\eeq
where we have $\lam_2^q \sim 10^{-7}$; we have denoted a typical phase by $\omega$.  For $m_2\sim 1$~TeV, we then find 
$d_q \sim 10^{-32} \sin \omega$~$e$ cm.  The current 90\% C.L. bound on neutron EDM is 
$d_n < 3.0 \times 10^{-26}$~$e$ cm \cite{Tanabashi:2018oca}, which indicates our model is not constrained much by the neutron EDM experiments. 

In order to go further and study electron EDM bound constraints, we need to have a measure of how 
large lepton flavor violating couplings can be in our model.  We will parametrize flavor violation 
by $\lam_a^{e\mu}$, $\lam_a^{\mu \tau}$, and $\lam_a^{e \tau}$, for tree-level transitions mediated by $H_a$ for $a=2,3$, in an obvious notation.  Since $H_1$ couplings to leptons are severely suppressed, we will only consider the dominant contributions from $H_a$ for $a=2,3$.

With the above assumptions, we have 
\beq
\Gamma(\ell \to 3\, f) \approx \frac{\lam_a^{f \,2} \lam_a^{f \ell \, 2}}{1536 \pi^3}
\frac{m_\ell^5}{m_a^4}\,,
\label{Gamellto3f}
\eeq
where $\ell = \mu, \tau$ and $f$ is a light final state charged lepton; we have ignored the 
effect of final state masses on the phase space.  

We have $\lam_a^e \sim 3 \times 10^{-4}$,  
$\lam_a^\mu \sim 6 \times 10^{-2}$, and as before $m_a\sim$~TeV.  We then find 
\beq
\Gamma(\mu \to 3 \,e) \sim 10^{-28} |\lam_a^{e\mu}|^2\, m_\mu
\label{muto3e}
\eeq
and 
\beq
\Gamma(\tau \to 3 \, \mu) \sim 10^{-18} |\lam_a^{\mu\tau}|^2 \,m_\tau
\label{tauto3mu}
\eeq
where $m_\mu $ and $m_\tau$ are the masses of the $\mu$ and $\tau$ leptons, respectively.  The width $\Gamma(\tau \to e \mu \mu)$ is given by the above formula, with 
$\lam_a^{\mu\tau}\to \lam_a^{e\tau}$.  The total widths are given by $\Gamma_\mu \approx 2.8 \times 10^{-18} \,m_\mu $ and $\Gamma_\tau \approx 1.3 \times 10^{-12} \,m_\tau$, 
in an obvious notation.  The current 90\% C.L. bounds on the above decays are 
BR$(\mu \to 3 \,e) < 1.0\times 10^{-12}$, 
BR$(\tau \to 3 \, \mu) < 2.1 \times 10^{-8}$, and BR$(\tau \to e \mu\mu) < 2.7 \times 10^{-8}$ \cite{Tanabashi:2018oca}.  Hence, we find 
\begin{eqnarray}
|\lam_a^{e\mu}| \lsim 0.2,\,\, |\lam_a^{\mu \tau}| \lsim 0.2,\,\,{\rm and}\,\, |\lam_a^{e \tau}| \lsim 0.2.  \label{ellto3f}
\end{eqnarray}

The dominant contribution to the electron EDM $d_e$, based on our model assumptions will then be 
mediated by a 1-loop $H_a$ diagram through the flavor-changing $e \mu$ or $e \tau$ coupling of $H_a$.  We then estimate a typical value by 
\begin{eqnarray}
d_e&\sim& \frac{e\, \lam_a^{e \ell\,2}\, m_\ell\, \sin \omega}{16\pi^2\, m_a^2}\\
&\sim& 
\begin{cases} 
10^{-23}\,|\lam^{e\mu}_a|^2\sin\omega\,e~{\rm cm}\quad{\rm for}\,\ell=\mu\\
10^{-22}\,|\lam^{e\tau}_a|^2\sin\omega\,e~{\rm cm}\quad{\rm for}\,\ell=\tau
\end{cases}
\label{de}
\end{eqnarray}
Note that the while we are using the 
same notation for the phase $\omega$ as before, it only is meant to denote a typical phase and 
is not assumed to have the same numerical value.  The 90\% C.L. bound 
$d_e < 1.1 \times 10^{-28}$~$e$ cm~\cite{Andreev:2018ayy} then implies bounds of
\begin{eqnarray}
&|\lam^{e\mu}_a|\sqrt{\sin\omega}\lesssim 3\times 10^{-3}\label{EDM:yemu}\\
&|\lam^{e\tau}_a|\sqrt{\sin\omega}\lesssim 1\times 10^{-3}.\label{EDM:yetau}
\end{eqnarray}

\item $\mu\rightarrow e\gamma$: This process provides a potentially severe constraint on models of 
new physics.  Here, with our preceding assumptions, we expect the main contribution to 
$\mu\rightarrow e\gamma$ to arise from the $\lam_a^\mu$ and $\lam_a^{e\mu}$, or $\lam_a^{\mu\tau}$ and $\lam_a^{e\tau}$ couplings at 1-loop order, depending on if the internal fermion is a muon or tau.  The resulting effective operator can be estimated by 
\beq
O \sim \frac{e\, m_\ell \lam_a^{\mu\ell} \, \lam_a^{e\ell}}{16 \pi^2\, m_a^2}\,
\bar \mu \,\sigma_{\mu\nu} e F^{\mu\nu}\,,
\label{muegamma}
\eeq
where $\sigma_{\mu\nu} = (i/2)[\gamma_\mu, \gamma_\nu]$ and $\ell=\mu,\tau$.  This dipole operator yields the branching fraction 
\beq
\text{Br}(\mu \to e \gamma) \sim 3\times 10^{-4}|\lam^{e\ell}_a\,\lam^{\mu\ell}_a|^2\left(\frac{m_\ell}{\rm GeV}\right)^2\,,
\label{Brmuegamma}
\eeq
which should be compared with the 90\% C.L. constraint $\text{Br}(\mu \to e \gamma)<
4.2 \times 10^{-13}$~\cite{TheMEG:2016wtm}.  The bounds on the flavor off-diagonal couplings are then
\begin{eqnarray}
&|\lam^{e\mu}_a|\lesssim 8\times 10^{-3}\label{mue:yemu}\\
&|\lam^{e\tau}_a \lam^{\mu\tau}_a|\lesssim 2\times 10^{-5}\,.\label{mue:yetau}
\end{eqnarray}
If the bound in Eq.~(\ref{EDM:yetau}) is saturated and $\sin\omega\sim 0.1$, we obtain $|\lam^{\mu\tau}_a|\lesssim 2 \times 10^{-2}$.

\item $(g-2)$: From the above discussion we can conclude that the dominant contribution to the 
muon anomalous magnetic moment $g_\mu-2$ will come from  the flavor-changing $H_a$-$\mu$-$\tau$ coupling $\lam_a^{\mu \tau}$ which is the least constrained.  We can then estimate the contribution to $(g_\mu-2)/2$ by 
\beq
\Delta a_\mu \sim \frac{\lam_a^{\mu\tau\,2} \,  m_\tau^2}{16\pi^2 \,{m_a^2}}\,,
\label{g-2}
\eeq
which yields $|\Delta a_\mu| \lesssim 2\times 10^{-12}$ (for $\sin\omega\sim 0.1$), which is too small to account for the current $\sim 3.5 \sigma$ 
anomaly \cite{Tanabashi:2018oca}.
\end{itemize}

Ref.~\cite{Babu:2018uik} suggests that a Yukawa flavor structure of $\lam^{ij}\sim {\rm min}(m_i,m_j)/v_{EW}$ is in good agreement with data\footnote{ Another well-known flavor structure is the Cheng-Sher~\cite{Cheng:1987rs} ansatz $\lam^{ij}=\sqrt{m_im_j}/v_{EW}$.  However, Ref.~\cite{Babu:2018uik} suggests that $\lam^{ij}\sim {\rm min}(m_i,m_j)/v_{EW}$ is in better agreement with observations}.  Here, we will determine the compatibility of our Yukawa couplings with this ansatz.  Note that since the leptons only obtain their mass from one Higgs doublet ($\Phi_2$) in the $v_3\rightarrow 0$ limit, their couplings to $\Phi_2$ will be diagonal after diagonalizing the lepton mass matrix.  Since $H_2$ is mostly $\Phi_2$, its couplings are also mostly diagonal while $H_3$ couplings can be flavor off-diagonal.  However, as mentioned above, the scalar mass eigenstates after EWSB are superpositions of the components of $H_{2,3}$ with a small component from $H_1$, and can have flavor off-diagonal couplings to leptons.  We maintain the generic notation $\lambda_a^{ij}$.

In order to keep $\lam_a^\tau\sim1$, for the charged leptons we modify the ansatz of Ref.~\cite{Babu:2018uik} to $\lam_a^{ij}\sim {\rm min}(m_i,m_j)/m_\tau$.    Hence, we have $\lam_a^{e\mu}\sim \lam^{e\tau}_a\sim 3\times 10^{-4}$, and $\lam^{\mu\tau}_a\sim 0.06$.  The constraint from $\ell\rightarrow 3f$ in Eq.~(\ref{ellto3f}) is clearly satisfied.  For $\sin\omega\sim 0.1$, the bounds in Eqs.~(\ref{EDM:yemu},\ref{EDM:yetau},\ref{mue:yemu},\ref{mue:yetau}) are also satisfied, although we are within order one of many of these bounds.  Hence we conclude that our mechanism is viable, in agreement with low energy observables, and if this ansatz for the charged lepton Yukawas holds we may expect to see a signal in the electron EDM or $\mu\rightarrow e\gamma$~\cite{Baldini:2018nnn}.

\begin{figure*}[tb]
\subfigure[]{\includegraphics[width=0.45\textwidth,clip]{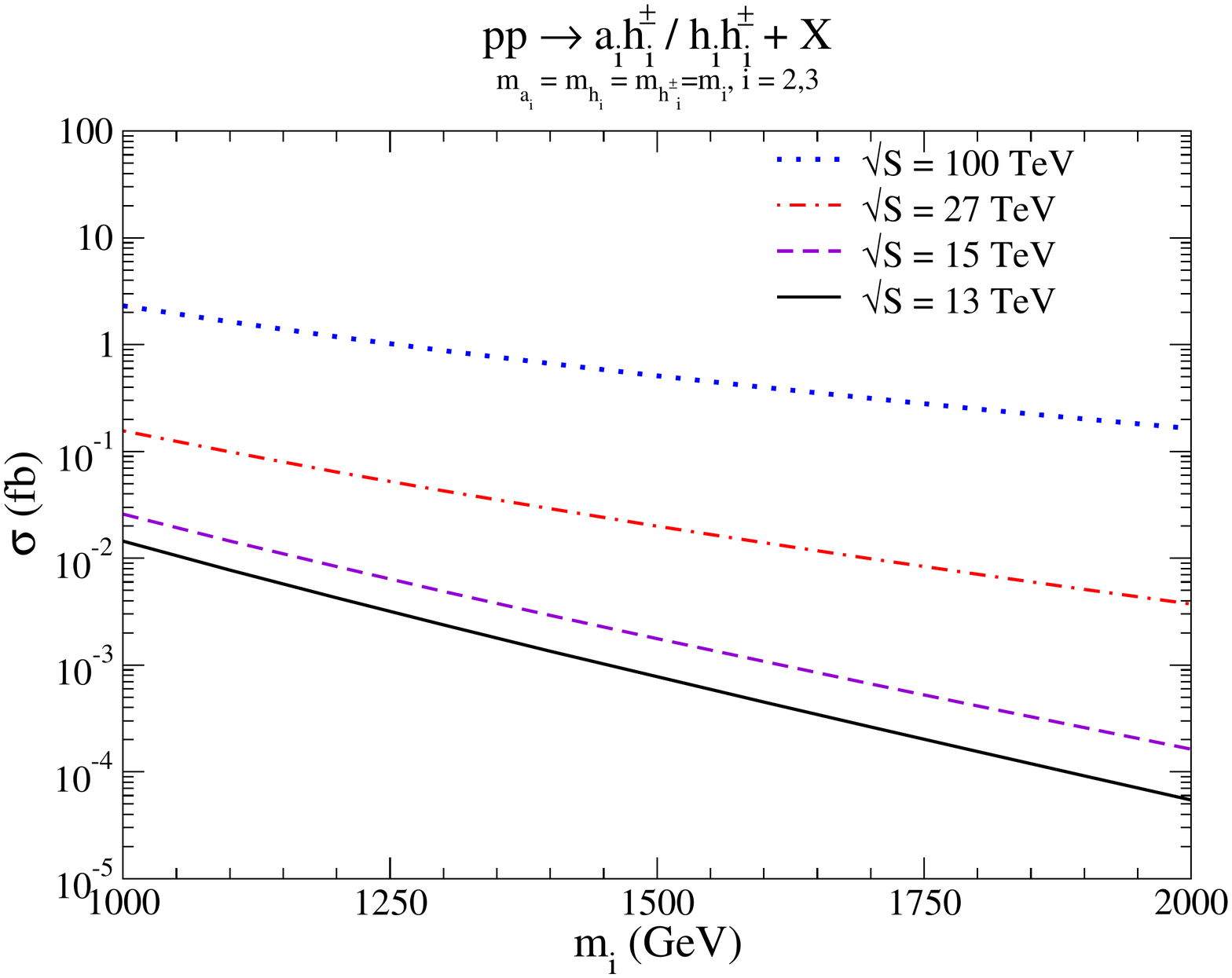}}
\subfigure[]{\includegraphics[width=0.45\textwidth,clip]{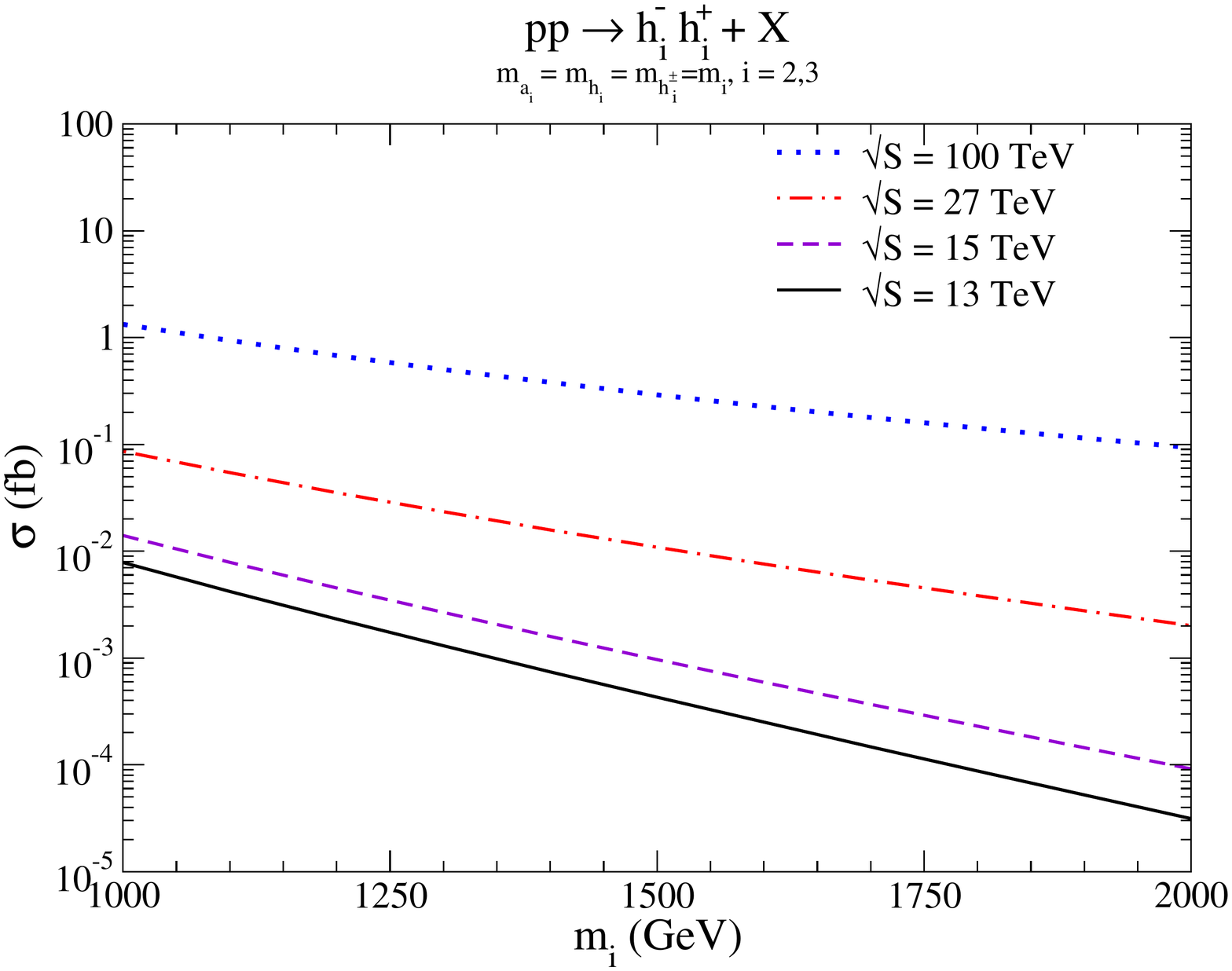}}
\caption{Production cross sections for heavy scalars (a) $h_ih_i^\pm$ and $h_ih_i^\pm$, and (b) $h_i^+h_i^-$.  Both Drell-Yan and VBF production mechanisms are included for all processes.  We show the cross sections for lab frame energies of (blue dotted) $\sqrt{S}=100$~TeV, (red dash-dot) $\sqrt{S}=27$~TeV, (violet dashed) $\sqrt{S}=15$~TeV, and (black solid) $\sqrt{S}=13$~TeV.}
\label{pairproduction}
\end{figure*}

\section{Collider Searches \label{sec:colliders}}

Now we discuss some of the aspects of the signals of our model at proton-proton colliders.  First, we concentrate on the pair production rates of the new heavy scalars.  After electroweak symmetry breaking, the two heavy Higgs doublets in the Higgs basis can be decomposed as
\begin{eqnarray}
H_i=\frac{1}{\sqrt{2}}\begin{pmatrix} \sqrt{2} h_i^\pm \\ h_i+i\,a_i\end{pmatrix},\quad{\rm for}~i=2,3.
\end{eqnarray}
Hence, we have 4 charged states, 2 pseudoscalar bosons, and 3 scalar bosons (including the scalar $h_1$ from $H_1$).  The Goldstone bosons completely reside within $H_1$.  Electroweak precision constraints generally require at least one of the neutral scalars $a_i,h_i$ to be mass degenerate with the charged scalars $h_i^\pm$~\cite{Barbieri:2006dq}.  Hence, for simplicity we will assume that $h_i^\pm,h_i,$~and~$a_i$ have a common mass $m_i$ for each $i=2,3$.  Production cross sections are computed in \texttt{MadGraph5\_aMC@NLO}~\cite{Alwall:2014hca} using a model generated via \texttt{FeynRules}~\cite{Alloul:2013bka}.

In Fig.~\ref{pairproduction} we show the pair production rates for various di-scalar final states: (a) $h_i h_i^\pm$ and $a_i h_i^\pm$, and (b) $h_i^+h_i^-$ for $i=2,3$.  We provide cross sections for (black solid) the $\sqrt{S}=13$~TeV LHC, (violet dashed) the proposed $\sqrt{S}=15$~TeV upgrade of the LHC~\cite{MedinaMedrano:2653736}, (red dot-dash) the proposed $\sqrt{S}=27$~TeV upgrade of the LHC (HE-LHC)~\cite{Abada:2019ono}, and the proposed $\sqrt{S}=100$~TeV colliders (FCC-hh/SppC)~\cite{CEPC-SPPCStudyGroup:2015csa,Benedikt:2018csr}.   The production cross sections for $h_ia_i$, although not shown, are within $\sim5-20\%$ of $h_i^+h_i^-$.  The production modes considered here depend almost exclusively on the gauge couplings of the heavy scalars, and hence have minimal dependence on the model parameters.  The di-scalar final states $h_ih_i$ and $a_ia_i$ will depend on trilinear scalar couplings and not gauge couplings, so we do not discuss them.  Finally, we have included both Drell-Yan and production in association with two jets (similar to vector boson fusion).  However, we find the production with two jets to be always subdominant.  This is in contrast to the SM case, where the vector boson fusion production rate of the Higgs boson competes with gluon fusion for Higgs mass $\gtrsim1$~TeV~\cite{deFlorian:2016spz}.

The benchmark luminosity for the 13 and 15 TeV LHC is 3 ab$^{-1}$, for the HE-LHC 15 ab$^{-1}$, and for FCC-hh/SppC 30 ab$^{-1}$.  Hence, for $m_i\sim 1-2$~TeV, we can expect between zero and 40 events at the high luminosity 13 TeV LHC.  At 15 TeV, the situation is slightly improved to an expected number of events between 1 and 80.  With between 30 and 2,300 events, the HE-LHC would be likely to be sensitive to much of the relevant parameter region and test our model.  Of course, the situation is most promising at the FCC-hh/SppC with between 2,800 and 50,000 events.  These predictions for the number of events are robust, since the production channels we consider are fully determined by gauge couplings.  While 40-80 events at the LHC may seem small, as we discuss below, the decays of these heavy scalars can be striking and with small background.  Hence, the LHC may be able to probe masses around $1$~TeV, while future colliders may be needed for masses at or above 2 TeV.  A full collider study would be necessary to determine the full reach of these machines.

We will now discuss the decays of the new scalars. Due to the vev hierarchy, from Eq.~(\ref{eqn:rotation}) the mixing between $\Phi_1$ and $\Phi_2$ is $v_2/v_1\sim 1\%$ and between $\Phi_1$ and $\Phi_3$ is much smaller as assumed before. The decays of the heavy scalars into quark, gauge boson, and di-Higgs channels depend on the mixing and are highly suppressed. Hence, the heavy scalars predominantly decay into leptons via their Yukawa couplings.  The neutral scalars $h_2$ and $a_2$ each decay mainly to a $\tau$ pair. Since we require $m_{3}\gg m_{R3}$ in our baryogenesis mechanism, the neutral scalars $h_3$ and $a_3$ each decay primarily to a heavy ($\nu_{R3}$) and a light neutrino and potentially similarly into charged leptons. For the charged scalars, since $H_2$ couples according to charged lepton masses, $h_2^\pm$ will decay to a $\tau$ and a light neutrino.  Since $H_3$ couplings are not necessarily as hierarchical as the charged fermions, $h_3^\pm$ can decay into $\mu,e$ and $\nu_{R3}$, as well as a $\tau$ and 
$\nu_{R3}$. 

With our sample parameters, used to derive \eq{eq:eps1}, only $\nu_{R3}$ is potentially accessible at collider experiments, 
with $\nu_{R1,2}$ being too heavy ($\sim 10$~TeV) to produce at the LHC and likely other 
envisioned facilities.  Here, assuming that $m_{R3}\gsim 100$~GeV, $\nu_{R3}$ can decay to SM gauge bosons via mixing,  to $H_1$ and a light neutrino through direct coupling leading to a ``Dirac'' mass of $m_{D3}\sim$~keV, or in three-body decays via an off-shell heavy scalar into leptons plus missing energy.  The mixing angle $\theta\sim m_{D3}/m_{R3}\sim 10^{-8}$ for $\nu_{R3}$-$\nu$ 
mixing leads to the following estimate  
\begin{eqnarray}
\Gamma(\nu_{R3}\sim W^\pm \ell^\mp)&\sim& 4 \Gamma(\nu_{R3}\to \nu_L \, Z)\nonumber\\ 
&\sim& \frac{\theta^2}{8\pi} \frac{m^3_{R3}}{v^2} \lsim 10^{-16} ~\text{GeV},
\label{nuV}
\end{eqnarray}
with $V=W,Z$.  We also find 
\beq
\Gamma(\nu_{R3}\to \nu_L \,h_1) \sim \frac{1}{32\pi}|\lambda_1^{\nu_{R3}}|^2 m_{R3} \lsim 10^{-16}~\text{GeV},
\label{nuh1}
\eeq
with $|\lambda_1^{\nu_{R3}}|\lsim 10^{-8}$ in our preceding example.  Finally, 
we also find, in analogy to Eq.~(\ref{Gamellto3f}),
\beq
\Gamma(\nu_{R3}\to \nu_L \,\ell \,\ell) \sim  \frac{|\lam_{2,3}^\ell \,\lam_{2,3}^\nu|^2}{1536 \pi^3}\frac{m_{R3}^5}
{m_{2,3}^4} \lsim 10^{-19}~\text{GeV}.
\label{nuellell}
\eeq

The above estimates imply that in our example the $\nu_{R3}$ decays  
would be quite displaced, on the order of meters.  This could in principle lead to very unique signals.  However, the proximity of the estimates (\ref{nuV}) and (\ref{nuh1}) suggests 
that a more careful study is needed to decide the dominant decay mode, but one could end up with 
similar rates for the first two possibilities.  Since the example parameters used to illustrate 
the viability of our baryogenesis mechanism were only one of many possible solutions, we do not 
offer a more detailed analysis here, but suffice it to say that the model can potentially 
yield interesting signals of $\nu_R$ decays.  

The phenomenology of SM-like Higgs boson, $h_1$, can also be altered.  Initially, in the Higgs basis of $H_1,H_2,H_3$, the coupling of $h_1$ are precisely the same as in the SM.  However, there can be mixing between neutral scalars $h_1$ and $h_2$ via quartic interactions in the Higgs potential.  For order one couplings, these mixings could be expected to be of the size $\sim v_1\,v_{2}/m_{2}^2$ which, assuming TeV scale heavy Higgses, is around $\sim 0.1\%$ for $h_2$.  Since the mixing with the heavy scalars are small, the production rate and main decay rates ($b\bar{b}$, $WW$, $ZZ$, $\gamma\gamma$) of $h_1$ are little changed.  However, the branching ratios into rarer modes, such as $\mu^-\mu^+$, can be altered.  The SM-like Yukawa coupling of $h_1$ to muons is $ m_\mu/v_{EW}\sim 4\times10^{-4}$, while the $h_2$ coupling to muons is $m_\mu/\sqrt{2}m_\tau\sim 0.04$.  Hence, after $0.1\%$ mixing with $h_2$, the coupling of $h_1$ to muons can be shifted from the SM by $\sim 10\%$.  The branching ratio of $h_1\rightarrow \mu^+\mu^-$ is then moved away from the SM value by $\sim 20\%$.  This shift is generically true of all charge leptons including $\tau$s.  While $h_1\rightarrow e^-e^+$ is unobservable at the LHC due to small electron couplings, this level of deviation in $h_1\rightarrow\mu^+\mu^-$ and $h_1\rightarrow \tau^+\tau^-$ will be observable at the high luminosity LHC with 3 ab$^{-1}$ or the HE-LHC with 15 ab$^{-1}$ of data~\cite{Cepeda:2019klc}.

\section{Summary}
In this paper we have presented a mechanism for the generation of the baryon asymmetry via heavy Higgs doublet decays into lepton doublets and right-handed neutrino singlets.  These decays produce an asymmetry in the lepton doublets that then gets processed into a baryon asymmetry via the electroweak sphalerons.  This scenario is a nearly minimal extension of the SM, in which we only need right-handed neutrinos which can help explain neutrino masses, and additional Higgs doublets.  Since the Yukawa couplings between the SM Higgs boson and neutrinos is constrained to be small,  at minimum two additional Higgs doublets are required to guarantee that the asymmetry parameter in Eq.~(\ref{eps-def}) is sufficiently large.

In addition to generating the baryon asymmetry, this scenario could have many signatures at current and future experiments.  To generate the baryon asymmetry, there needs to be a misalignment between the Yukawas of the different Higgs doublets.  Once all Higgs doublets obtain a vev, this necessarily leads to flavor changing currents in the lepton sector as well as EDMs.  As shown above, the baryon asymmetry can be generated and current constraints on charged lepton flavor violation accommodated within a realistic Yukawa structure.  Furthermore, future $\mu\rightarrow e\gamma$ and electron EDM experiments may be expected to show signatures of this baryon asymmetry mechanism.

Finally, we studied the collider signatures of the heavy Higgs doublets.  Via di-scalar production, the scenario presented here can provide striking signatures of many leptons, missing energy, $b$-jets, and possibly displaced vertices.  While the di-scalar production rates can be favorable at the LHC, future colliders may be needed to observe much of the interesting parameter space.    Additionally, we may expect the observed Higgs boson decays into muons and taus, $h_1\rightarrow \mu^+\mu^-/\tau^+\tau^-$, to differ from SM predictions by upwards of $20\%$.  This is an observable amount of deviation at the high luminosity LHC with 3 ab$^{-1}$ or the HE-LHC with 15 ab$^{-1}$ of data~\cite{Cepeda:2019klc}.

\textbf{Acknowledgments}  I.M.L. would like to thank the Institute for Theoretical Physics at Universit{\"a}t Heidelberg for their hospitality during the completion of this work and Prof. Dr. Tilman Plehn for insightful commentary on the nature of BSM models.  H.D. is supported by the United States Department of Energy under Grant Contract DE-SC0012704.  I.M.L. is supported in part by the United States Department of Energy grant number DE-SC0019474.  M.S. is supported in part by the State of Kansas EPSCoR grant program, United States Department of Energy grant number DE-SC0019474, and a Summer Research Scholarship from the University of Kansas.   The data to reproduce the plots has been uploaded with the arXiv submission or is available upon request.

\bibliographystyle{myutphys}
\bibliography{troika-refs}
\end{document}